\documentclass[twocolumn,10pt]{tsfp8}
\usepackage{flushend}
\usepackage{graphicx}
\usepackage[authoryear,round]{natbib}

\title{Turbulence transition in the asymptotic suction boundary layer}

\renewcommand{\expauthors}{1.1}

\author{Tobias Kreilos
    \affiliation{
      Fachbereich Physik \\
      Philipps-Universit\"at Marburg\\
      Renthof 6, D-35032 Marburg, Germany\\
      tobias.kreilos@physik.uni-marburg.de
    }	
}

\author{Taras Khapko
  \affiliation{Linn{\'e} FLOW Centre \\
              KTH Mechanics \\
              Osquars Backe 18, SE-100 44 Stockholm, Sweden \\
              taras@mech.kth.se}
}

\author{Tobias M. Schneider
  \affiliation{Max Planck Institute for Dynamics and Self-Organization \\
              Am Fassberg 17, D-37077 G\"ottingen, Germany\\
              tobias.schneider@ds.mpg.de}
}
\author{Gregor Veble
  \affiliation{Pipistrel d.o.o. Ajdovščina \\
              Goriška c. 50a, SI-5270 Ajdovščina, Slovenia \\
              gregor.veble@ung.si}
}
\author{Yohann Duguet
  \affiliation{LIMSI-CNRS \\
              UPR 3251, F-91403 Orsay, France \\
              duguet@limsi.fr}
}
\author{Philipp Schlatter
  \affiliation{Linn{\'e} FLOW Centre \\
              KTH Mechanics \\
              Osquars Backe 18, SE-100 44 Stockholm, Sweden \\
              pschlatt@mech.kth.se}
}
\author{Dan S. Henningson
  \affiliation{Linn{\'e} FLOW Centre \\
              KTH Mechanics \\
              Osquars Backe 18, SE-100 44 Stockholm, Sweden \\
              henning@mech.kth.se}
}
\author{Bruno Eckhardt
  \affiliation{Fachbereich Physik \\
              Philipps-Universit\"at Marburg \\
              Renthof 6, D-35032 Marburg, Germany \\
              bruno.eckhardt@physik.uni-marburg.de}
}

\newcommand{\ecf}{E_{cf}}
\newcommand{\ecft}{\langle E_{cf}\rangle_T}

\begin{document}

\maketitle   

\fontsize{9}{11}\selectfont

\section*{ABSTRACT}
We study the transition to turbulence in the asymptotic suction boundary layer (ASBL)
by direct numerical simulation. Tracking the motion of 
trajectories intermediate between laminar and turbulent states 
we can identify the invariant object inside the laminar-turbulent boundary, the edge state. 
In small domains, the flow behaves like a travelling wave over short time intervals.
On longer times one notes that the energy shows strong bursts at regular time intervals.
During the bursts the streak structure is lost, but it reforms, translated in the spanwise
direction by half the domain size.
Varying the suction velocity allows to embed the flow into a family of flows that interpolate between 
plane Couette flow and the ASBL.
Near the plane Couette limit, the edge state is a travelling wave. Increasing the suction, the travelling wave and a symmetry-related copy of it
undergo a saddle-node infinite-period (SNIPER) bifurcation that leads to bursting and discrete-symmetry shifts.
In wider domains, the structures localize in the spanwise direction, and the flow in the active region is similar to the one in small domains.
There are still periodic bursts at which the flow structures are shifted, but the shift-distance is no longer connected to a discrete symmetry of the flow geometry.
Two different states are found by edge tracking techniques, one where structures are shifted to the same side at every burst and one where they are alternatingly shifted to the left and to the right.

\section*{Introduction}
Transition to turbulence in shear flows has puzzled physicists for decades. 
In many systems the transition occurs while the laminar flow profile is linearly stable \citep{Grossmann2000}.
The transition is connected to the appearance of coherent three-dimensional invariant structures, as identified for example in
plane Couette flow \citep{Nagata1990,Clever1997,Kawahara2001,Wang2007,Viswanath2007,Gibson2008a,Halcrow2009}
and most prominently pipe flow \citep{Faisst2003,Wedin2004,Willis2008},
where flow structures very similar to the numerically calculated ones have been observed experimentally by \citet{Hof2004}.
Edge states are invariant attractors inside the laminar turbulent boundary. Through their codimension-one stable manifold,
they guide the transition to and the decay from turbulence \citep{Toh2003,Skufca2006,Schneider2007b,Eckhardt2008,Schneider2009,Vollmer2009,Lebovitz2009,Kreilos2012}
and they are, at least in simple and confined geometries, transiently visited by turbulent trajectories \citep{Kawahara2005a,Schneider2007a}.

While most of the investigations on edge states have focused on internal flows, many interesting flow situations are external flows.
Attempts to characterize the edge in the Blasius boundary layer \citep{Duguet2012,Cherubini2011} have to deal 
with the conceptual difficulties associated with the spatial development.
In two recent studies (\citet{Khapko2013} and \citet{Kreilos2013}) edge states in the asymptotic suction boundary layer have been documented.
In this work, we will summarize the results from those two studies and 
discuss additional aspects 

\section*{System and numerics}
The asymptotic suction boundary layer (ASBL) forms if fluid streams over a flat plate into which it is sucked with a constant, homogeneous suction velocity $V_s$, \citet{Hocking1975,Fransson2001}.
The laminar profile is translationally invariant, which makes this flow far easier to deal with in numerical simulations than a spatially growing boundary layer.
With the free-stream velocity $U_\infty$ parallel to the x-direction, the dynamic viscosity $\nu$, the definition $\delta = \nu/V_s$ and $y$ the wall-normal direction, the laminar flow profile reads:
\begin{equation}
 U_0(y) = (U_\infty(1-\mathrm e^{-y/\delta}), -V_s, 0).
\end{equation}
We base the Reynolds number on the boundary layer thickness $\delta$, $Re = \frac {U_\infty \delta}{\nu}$.
In a finite computational domain the velocity at the upper wall is imposed to be $U_\infty$.

We investigate the transition to turbulence in the ASBL by direct numerical simulations, using two pseudo-spectral codes, channelflow \citep{channelflow} and SIMSON \citep{simson}.
The codes employ periodic boundary conditions in the streamwise and spanwise directions and no-slip boundary conditions at the walls.

\section*{Edge states in small periodic flow domains}

\begin{figure}
 \centering
 \includegraphics[width=\linewidth]{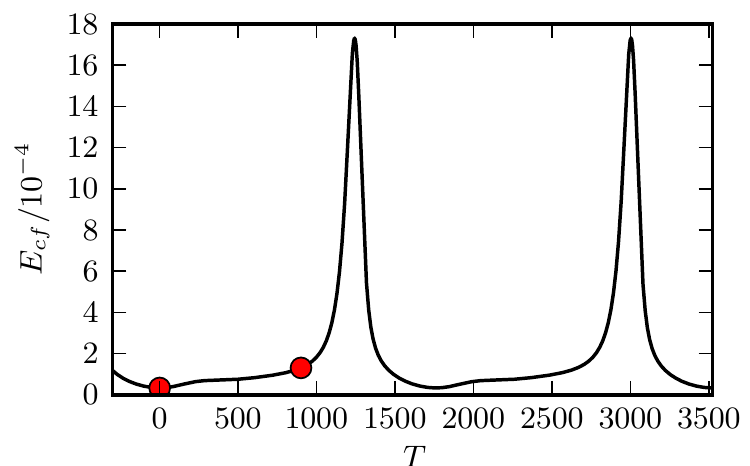}
 \caption{\label{fig:energyH}
  The cross-flow energy $\ecf$ for the edge state in a small periodic domain of size $L_x\times L_z=6\pi\times3\pi$ at Reynolds number $500$.
  The time evolution consists of long calm period, where the cross-flow energy is almost constant and short energetic bursts at regular intervals.
  While we only plot two peaks, the curve continues periodically.
  The two red circles mark the times where the snapshots in figure~\ref{fig:small_es} were taken.
 }
\end{figure}
\begin{figure}
  a)\vspace*{-1em}\\
  \hspace*{1em}
  \includegraphics[width=.95\linewidth]{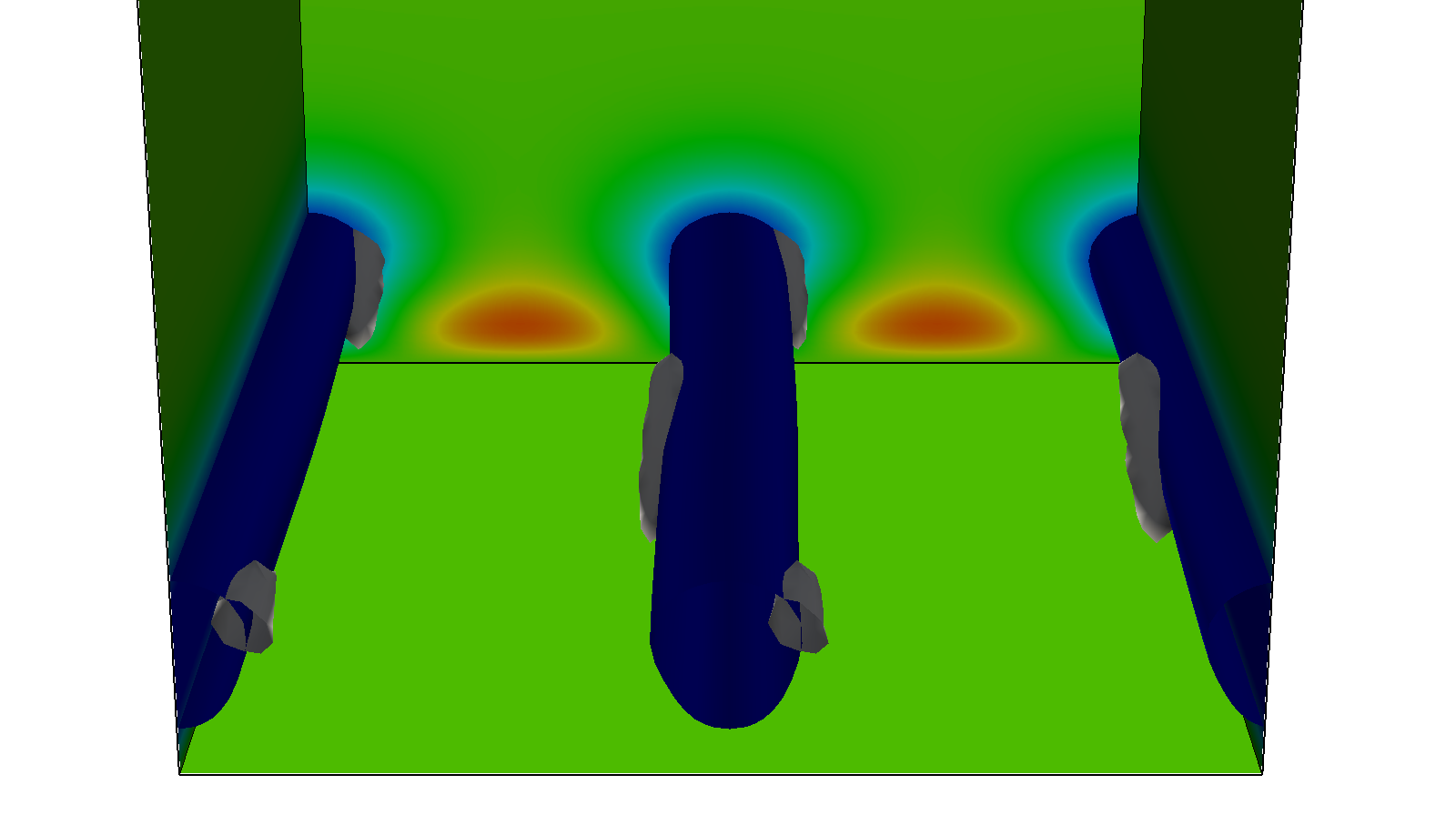} \\
  b)\vspace*{-1em}\\
  \hspace*{1em}
 \includegraphics[width=.95\linewidth]{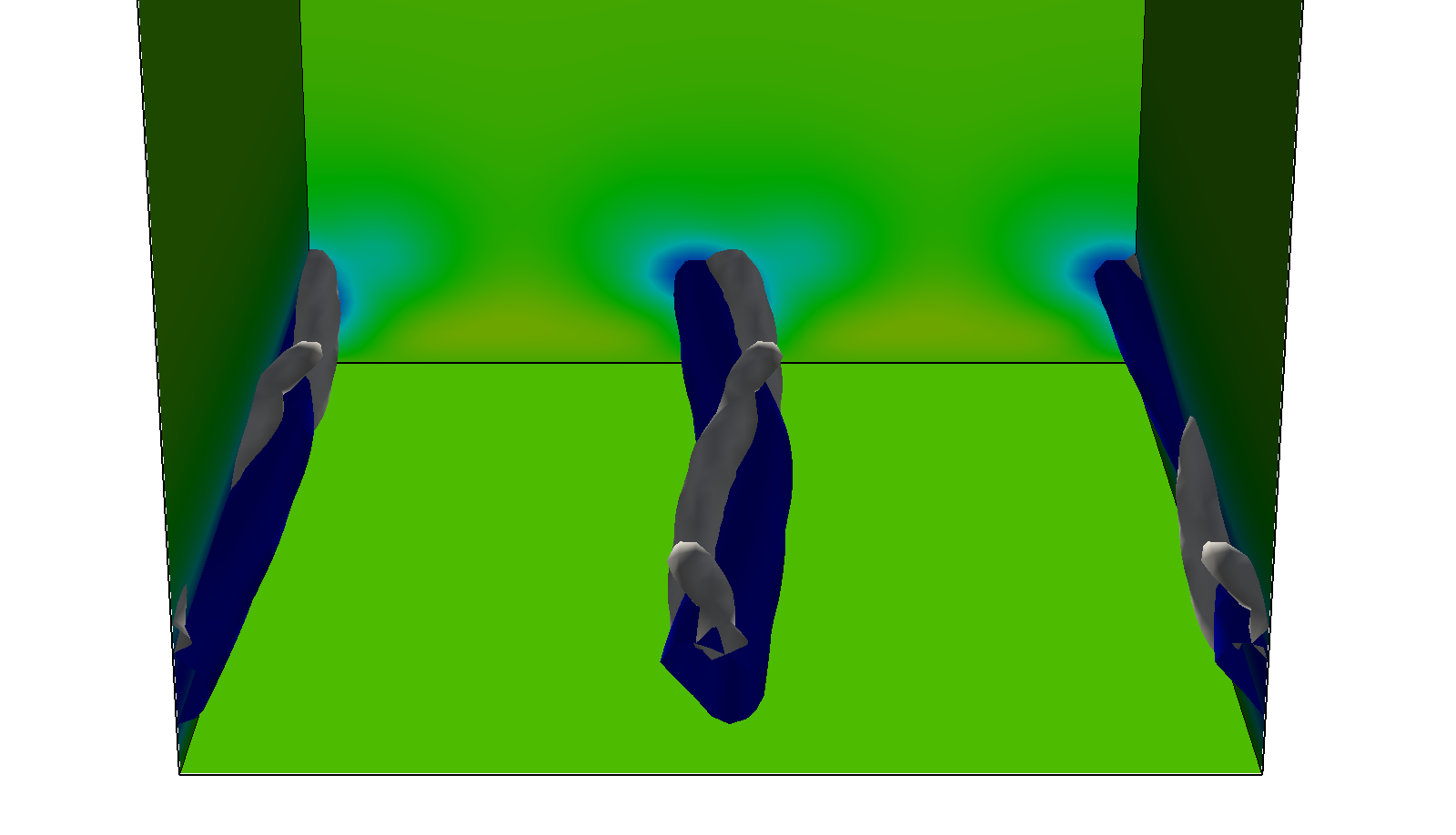}
 \caption{\label{fig:small_es}
  Snapshots of the edge state at two different times.
  We plot low speed streaks using isocontours of the downstream velocity perturbations $u=-0.25$ and visualize vortices by isocontours of the $\lambda_2$ vortex detection criterion.
  At the walls of the box, we color-code the downstream velocity perturbations.
  The box is shown twice in the spanwise direction, to help visualizing across the periodic boundaries, and we do not show the empty upper part of the box.
  a) At time $t=0$, $\ecf$ is minimal. There is one low- and one high-speed streak, accompanied by a pair of counter-rotating streamwise-oriented vortices. The gray isocontours are for $\lambda_2=-1\cdot10^{-4}$.
  The isocontours of the downstream velocity show that the streaks are almost straight.
  b) At time $t=900$, just before a burst. The streaks are a lot more contorted, the two vortices lean across the low-speed streak and are about to tear it apart. The isocontours are for $\lambda_2 = -5\cdot10^{-5}$.
 }
\end{figure}

In this section, we discuss the properties of the edge state in a domain of size $L_x \times L_y \times L_z = 6\pi \times25\times3\pi$ at Reynolds number $Re=500$; the resolution is $N_x\times N_y\times N_z = 96\times193\times96$.
It was shown by \citet{Kreilos2013} that in small periodic domains, the flow behaves like a travelling wave dominated by ordered modulated streaks over short time intervals.
On longer times one notes that the energy is not constant, as it would be for a 
travelling wave, but shows strong bursts with a period of $T=1760$.
The evolution of the cross-flow energy $\ecf = \frac 1{L_x L_z} \int_V (v^2+w^2) \mathrm dV$ is shown in figure~\ref{fig:energyH}, where the long calm phases and the short violent bursts can be seen.

Flow visualizations at the indicated points are shown in figure~\ref{fig:small_es}.
Isocontours of the fluctuating downstream velocity $u=-0.25$ are plotted in blue and indicate the location of low-speed streaks; $u$ is also color-coded at the walls of the box.
Vortices are visualized using the $\lambda_2$ vortex detection criterion \citep{Jeong1995}.
The box width is doubled in the spanwise direction, to facilitate visualization of the dynamics across the periodic boundaries.
\begin{figure*}
 \centering
 \includegraphics[width=.76\linewidth]{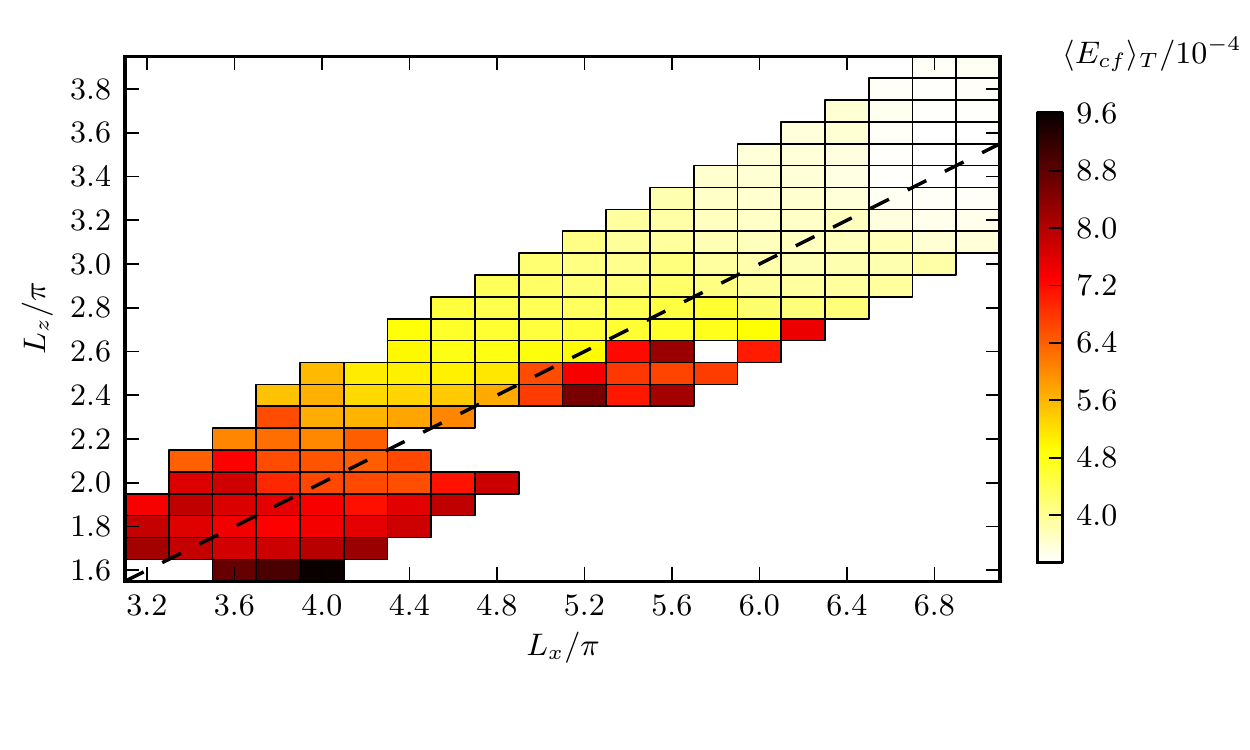}
 \vspace*{-3em}
 \caption{\label{fig:bs_ecf}
  Time-averaged cross-flow energy for different box sizes. Overall, the cross flow energy is smaller in larger boxes, but there are some notable exceptions to that general trend around $5.4\pi\times2.5\pi$.
  The dashed line indicates $L_x=2L_z$.
 }
\end{figure*}

\begin{figure*}
 \centering
 \includegraphics[width=.76\linewidth]{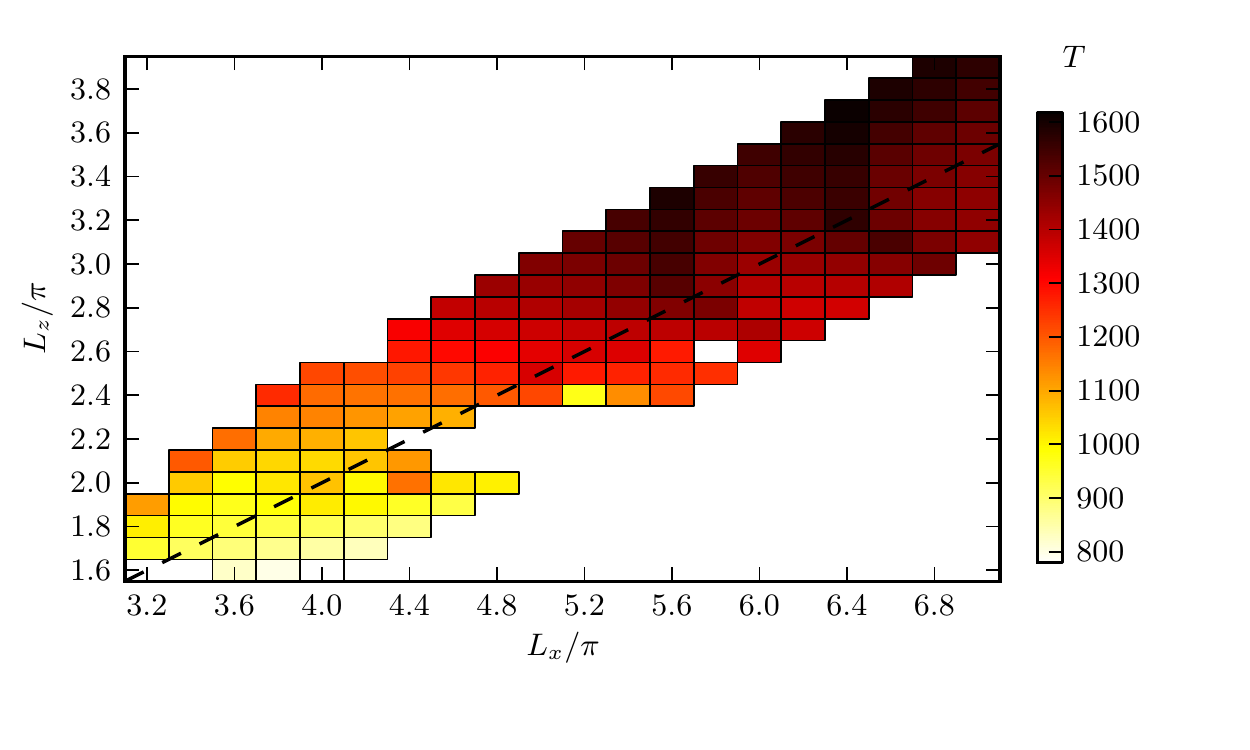}
 \vspace*{-3em}
 \caption{\label{fig:bs_t}
  Dependence of burst period $T$ on the streamwise and spanwise domain size.
  $T$ increases to the upper right corner, i.e.\ for larger boxes.
  Furthermore, $T$ is varies less along horizontal lines and increases more rapidly along the vertical axis for wider boxes.
  Boxes where no calculations have been performed or showed no periodic results are left blank.
  The dashed line indicates $L_x=2L_z$.
 }
\end{figure*}

The flow is composed of one low- and one high-speed streak and a pair of counter-rotating streamwise-elongated vortices.
During the calm phase, the streaks are almost straight.
The two vortices sustain the streaks by pushing high-speed fluid from the outer layers towards the wall, creating a high-speed streak, and lifting low-speed fluid from the wall into the main flow, creating the low-speed streak.
As time progresses, the streaks begin to tilt and the vortices lean over the low-speed streak, as can be seen in figure~\ref{fig:small_es} (b).
Finally, the streak is torn into two parts by the action of the vortices, the event corresponding to the peak in the cross-flow energy.
The two vortices then switch their positions and create a new pair of streaks, after which the process starts anew, but with all structures shifted by exactly half a box-width.
The state obeys a shift-and-reflect symmetry, i.e.\ it is invariant under a shift by $L_x/2$ followed by an inversion in $z$; the symmetry has not been imposed in the calculations but it is found a posteriori.
It is this symmetry that fixes the distance of the jumps to be exactly $L_z/2$, since only then there is no preferred direction and jumps to the left and the right are the same.

Varying the suction velocity allows to embed the flow into a family of flows that interpolate between 
plane Couette flow and the ASBL.
Near pCf, the edge state is a travelling wave. Increasing the suction, the travelling wave and a symmetry-related copy of it
undergo a saddle-node infinite-period (SNIPER) bifurcation that leads to the bursting and the discrete-symmetry shifts \citep{Kreilos2013}.

\section*{Variation of box size}

\begin{figure*}
 \centering
 \includegraphics[width=0.6\linewidth]{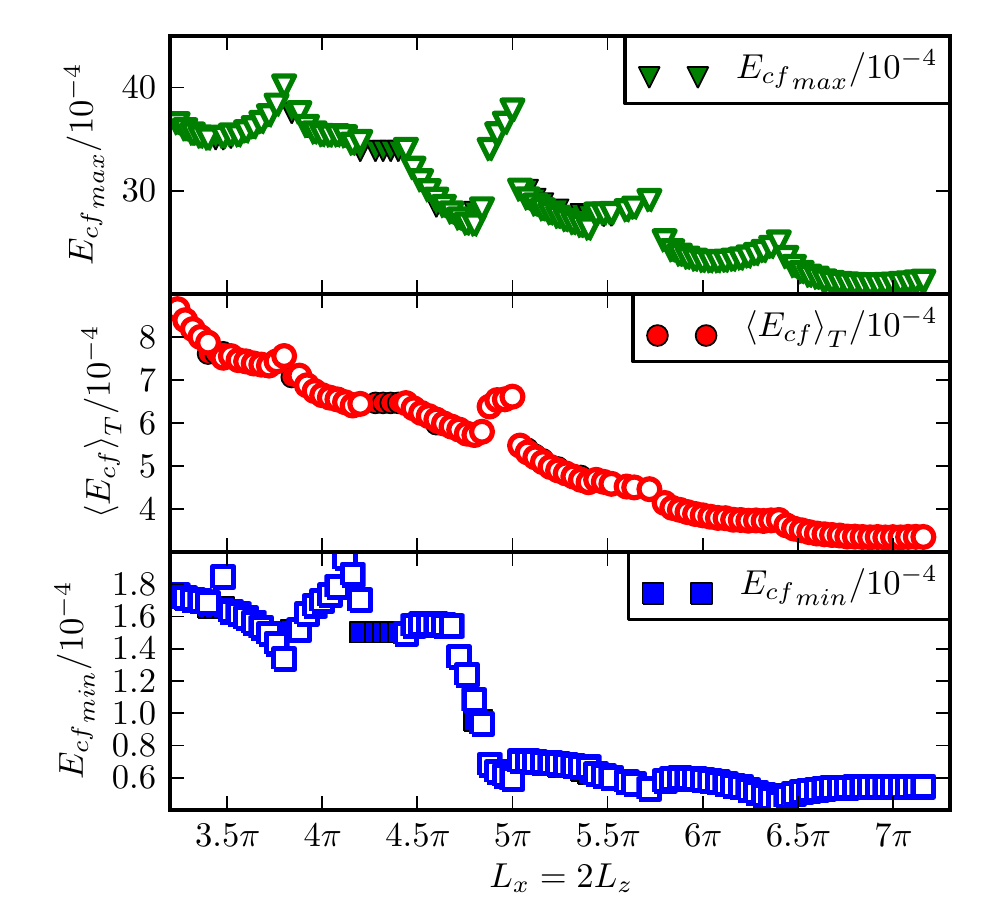}
 \caption{\label{fig:diag_ecf}
  The cross-flow energy along the ``diagonal'' $L_x=2L_z$.
  Upper panel: maximum of $E_{cf}$ during bursts.
  Middle panel: $\ecf$ averaged over one period.
  Lower panel: minimum of $\ecf$ after a burst.
  All calculations performed at $Re=400$, calculations where a shift-and-reflect symmetry was imposed are marked with full symbols.
  As a general trend, all of the three energies are lower in larger boxes.
  The minimal $\ecf$ is almost constant from $L_x=5.5\pi$ to the largest calculated boxes.
  Some resonance phenomenon due to the constraint from the periodic boundaries is suggested by the regular spacing of the extrema in the energies.
 }
\end{figure*}

 The periodicity in the streamwise and spanwise direction is an artificial constraint.
 In this section we study the dependence of the edge state on the numerical  parameters $L_x$ and $L_z$, using a Reynolds number of $Re=400$ and a fixed height of $L_y=10$, as in \citet{Kreilos2013}.
 In pCf, a box size close to $L_x = 2L_z$ has been found to be optimal for coherent structures \citep{Clever1997,Wang2007}.
 We hence focus on this ratio, and vary $L_z$ from $L_x/2-0.5\pi$ to $L_x/2+0.4\pi$ in steps of $0.1\pi$, with $L_x$ ranging from $3.2\pi$ to $7\pi$.
 In some boxes, edge state tracking did not converge to a periodic state within our integration time and we found a chaotic edge trajectory; in the following we only consider periodic states.
 The main result is that the general structure of all edge states remains the same: there are long calm phases, interrupted by bursts at regular time intervals. At every burst, the flow structures shift by half a box width.
 
 In order to characterize the states further, we study the dependence of cross-flow energy and burst period on domain size.
 We show them in figures~\ref{fig:bs_ecf} and \ref{fig:bs_t} as pseudocolor plots with $L_x$ on the horizontal axis and $L_z$ on the vertical one.
 The result for the box $L_x\times L_z$ is representad in the rectangle $(L_x\pm0.1\pi)\times(L_z\pm0.05\pi)$. Rectangles where no calculations were performed or showed no regular behaviour are left white. The diagonal $L_x=2L_z$ is indicated by the dashed black line.
 
  Figure~\ref{fig:bs_ecf} shows the time-averaged cross-flow energy. 
  As a general trend, the cross flow energy is high in the lower left corner of the plot where the box size is small and low in the upper right corner, in larger boxes.
  The flow structures seem to have a preferred size and do not grow arbitrarily as the box size is increased.
  A notable exception from the overall trend are $13$ states in a small region around $5.4\pi\times2.5\pi$. 

  The bursting period $T$, presented in figure~\ref{fig:bs_t}, varies inversely to the cross-flow energy. It is short for small boxes and long for bigger boxes. 
  There are probably at least two important contributions to this effect.
  First, higher energies trigger the bursts more frequently. 
  The anti-correlation is easily seen by comparing figures~\ref{fig:bs_ecf} and \ref{fig:bs_t}.
  But a second contribution is revealed by a closer inspection of figure~\ref{fig:bs_t}: the bursting period along a horizontal line of constant box width is roughly constant, but it increases drastically along vertical lines, as the box width grows.
  In wider boxes, the oscillations of the streaks have more time to build up and the vortices can tilt more before they cross and the streak breaks up.
  
  We present a more detailed study along the diagonal $L_x=2L_z$ in figure~\ref{fig:diag_ecf}, where three different quantities based on the cross-flow energy are shown.
  In the upper panel, we show the maximum of $\ecf$ during a burst; it measures the intensity of a burst.
  In the middle panel, $\ecf$ is averaged over one period. And in the lower panel, we show the minimum of $\ecf$ after a burst.
  For all three quantities, there is a general trend to decrease for increasing box sizes.
  But the dependence is non-monotonic, especially ${\ecf}_{max}$ shows a drastic increase around $L_x=4.8\pi$, while ${\ecf}_{min}$ drastcially drops at the same time.
  Particularly interesting is the fact that there seems to be a rather regular variation, with a sharp maximum of $\ecft$ at rather regular spacing and a smooth minimum between.
  This hints at some resonance phenomenon due to the constraint from the periodic boundaries.

\section*{Spanwise localized edge states}

In the last section we have shown that in small periodic domains the edge tracking algorithm converges to qualitatively similar states for a wide range of box sizes.
If the spanwise domain size is increased to larger values, the spacing between the streaks becomes too large and structures will localize.
Edge states in such wide boxes have been studied in \citet{Khapko2013}.
In this work, we use a box of size $L_x\times L_y\times L_z = 3\pi\times25\times50$ with a resolution of $48\times193\times256$.
Interestingly, depending on the initial condition, the edge state tracking algorithm converges to three different states (up to trivial translations), which can be categorized by the direction of propagation.
All of the states again consist of longer calm phases and violent bursts where the flow structures break up and reform at a different spatial location.

\begin{figure}
  a)\vspace*{-1em}\\
  \hspace*{1em}
 \includegraphics[width=.95\linewidth]{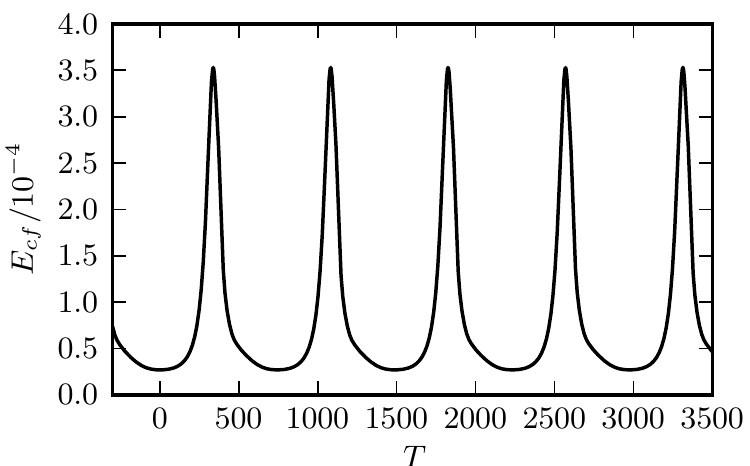}\\
   b)\vspace*{-1em}\\
  \hspace*{1em}
 \includegraphics[width=.95\linewidth]{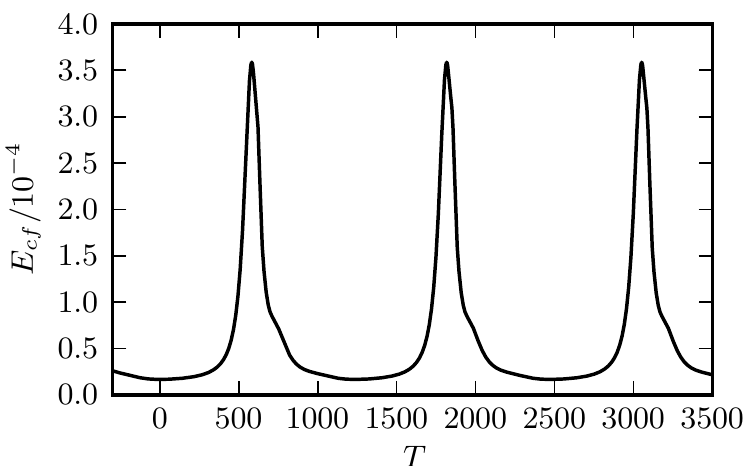}
 \caption{\label{fig:ecf_big}
  Cross-flow energy for localized edge states in a wide box at $Re=500$. 
  a) LR-state: at each burst the structures are displaced alternatingly to the left and the right. The interval between the jumps is $744$ time units, .
  b) L-state: at each burst the structures are displaced to the left. The interval between the jumps is $1234$ time units.
 }
\end{figure}

We found one state where the structures alternatingly jump to the left and the right, the cross-flow energy of the state is shown in figure~\ref{fig:ecf_big}(a); we name this state LR in the following.
Other initial conditions converged to a state which always jumps in the same direction, the cross-flow energy is shown in figure~\ref{fig:ecf_big}(b).
We have found the state that always jumps left (L) and the one that always jumps right (R).
The burst period of the states is $744$ for LR and $1234$ for L and R.
It is interesting to notice that these periods are both smaller than the period of the state in the small box, where it is $1760$.
While for increasing box sizes the period for the extended structures increased even further (figure~\ref{fig:bs_t}), it is smaller for the localized cases.

The mechanism for the streak breakup is similar to the one discussed above for the small boxes: after a burst, the streaks are mostly aligned downstream. One low speed streak is accompanied by a pair of vortices, which start to lean over the streak, cross and switch their positions in breaking up the streak.
While the physical origin of the bursts is still an instability of the low speed streak, the phase space structure must be different compared to the small box case.
The SNIPER-bifurcation in the original form described in \citet{Kreilos2013} requires periodic boundaries and the equivalence of the jump right and left.

\begin{figure*}
 \centering
 \includegraphics[width=\linewidth]{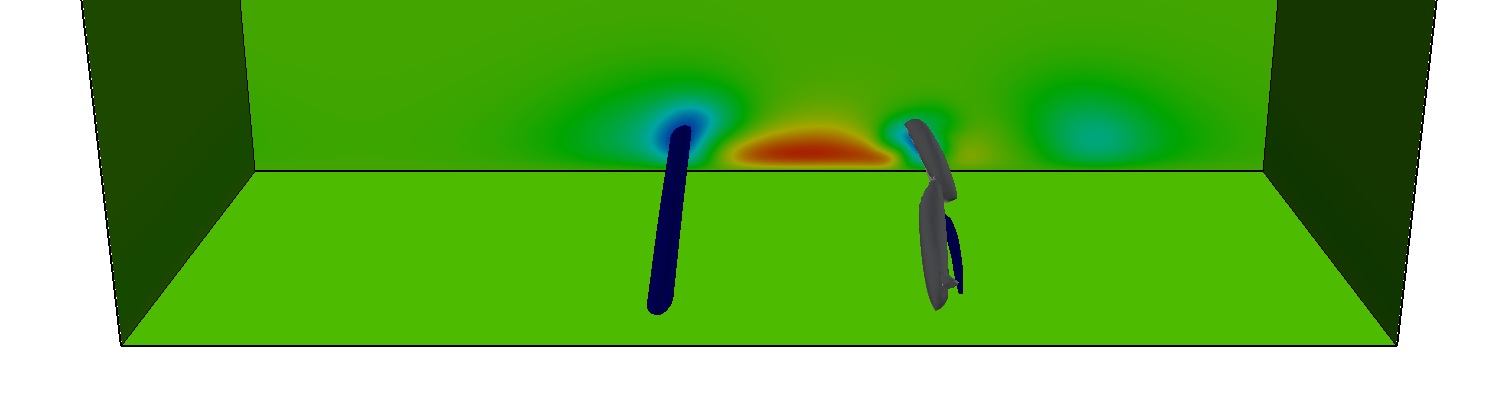}
 \includegraphics[width=\linewidth]{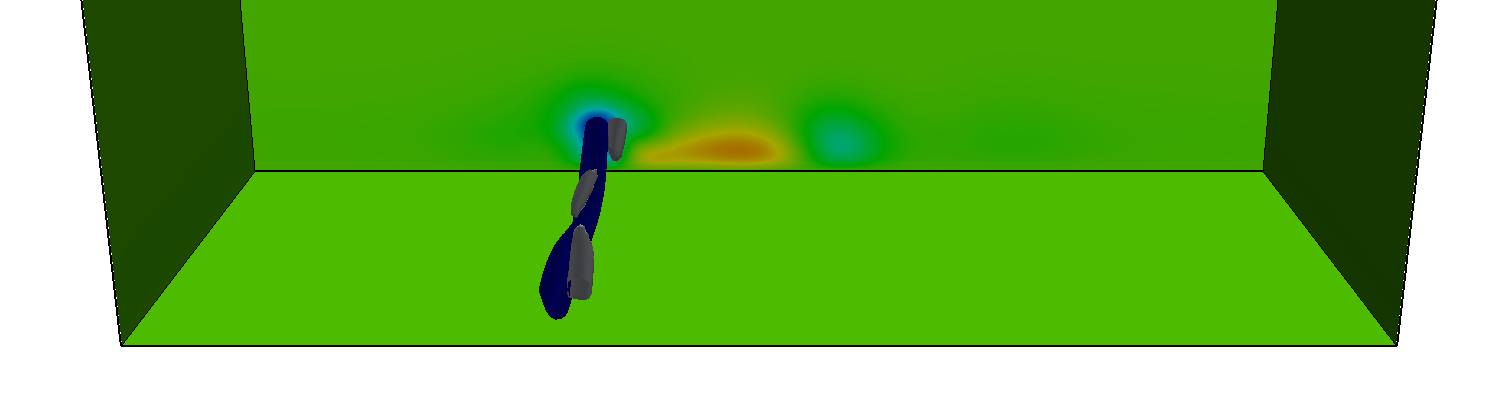}
 \caption{\label{fig:es_wide}
  Visualizations of the edge states in wide boxes.
  We show the isocontour $u=-0.33$ in blue and visualize vortices by the isocontour $\lambda_2 = -0.0008$.
  On the back of the box, the downstream perturbations are color-coded.
  a) The upper panel shows the LR-state at the minimum of $\ecf$. There is a low-speed streak on the left, followed by a strong high-speed streak and a low speed streak covered by two vortices. In due course, the vortices will tear this streak apart.
  b) The L-state at the minimum of $\ecf$. There is only one strong low-speed streak. Again, we see one pair of counter-rotating vortices, leaning over the low-speed streak.
 }
\end{figure*}

\section*{Conclusions}

We have studied edge states in the asymptotic suction boundary layer, both in small domains close to a minimal flow unit \citep{Kreilos2013} and in spanwise extended domains \citep{Khapko2013}, where the flow structures localize.
The localization of edge states in larger domains has also been observed in plane Couette flow \citep{Schneider2010a,Duguet2009} and pipe flow \citep{Mellibovsky2009}
In pCf it has been possible to identify a snaking bifurcation responsible for the localization, see \citet{Schneider2010}.

In the ASBL, for all cases where the edge state tracking converged to a non-chaotic state, the general structure and dynamics of the state are the same.
The flow exhibits long quiescent phases, interrupted by regular violent bursts.
During the quiet phase, the dynamically active part of the flow consists of a low speed streak which is flanked by a pair of counter-rotating vortices.
These vortices destabilize the streak and, while originally almost aligned downstream, start to tilt and lean over the streak.
As the vortices overlap, they break up the streak and a turbulent burst occurs.
Afterwards, the structures reform at a different spanwise location.
This location is dictated by symmetry in the small domain, where structures are translated by exactly half a box width.
In the wide domains, we found states where the structures alternatingly are displaced left and right and states where they are always shifted in the same direction.

In a small periodic domain, the flow pattern is robust under variations of the size of the computational domain.
We have shown that the time interval between two consecutive bursts grows as the box size is increased, while the time-averaged cross-flow energy decreases.
If the ratio of streamwise and spanwise domain size is kept fixed at two, some resonances due to the imposed wavelenghts can be observed.

Similar recurrence patterns as described here have also been suggested to exist in edge trajectories in spatially developing boundary layers as discussed by \citet{Duguet2012}.
It remains an open question, how the three edge states in the wide domain are connected and whether their origin can be traced to something similar like a SNIPER-bifurcation.

\bibliographystyle{tsfp8}

\begin{thebibliography}{35}
\expandafter\ifx\csname natexlab\endcsname\relax\def\natexlab#1{#1}\fi

\bibitem[Cherubini {\em et~al.\/}(2011)Cherubini, {De Palma}, Robinet \&
  Bottaro]{Cherubini2011}
Cherubini, S, {De Palma}, P, Robinet, J-Ch \& Bottaro, A 2011 {Edge states in a
  boundary layer}. {\em Phys Fluids\/} {\bf 23}, 051705.

\bibitem[Chevalier {\em et~al.\/}(2007)Chevalier, Schlatter, Lundbladh \&
  Henningson]{simson}
Chevalier, M, Schlatter, P, Lundbladh, A \& Henningson, D~S 2007 {A
  pseudo-spectral solver for incompressible boundary layer flows}. {\em Tech.
  Rep.\/}. KTH Mechanics, Stockholm, Sweden.

\bibitem[Clever \& Busse(1997)]{Clever1997}
Clever, R~M \& Busse, F~H 1997 {Tertiary and quaternary solutions for plane
  {Couette} flow}. {\em J Fluid Mech\/} {\bf 344}, 137--153.

\bibitem[Duguet {\em et~al.\/}(2009)Duguet, Schlatter \&
  Henningson]{Duguet2009}
Duguet, Y, Schlatter, P \& Henningson, D~S 2009 {Localized edge states in plane
  {Couette} flow}. {\em Phys Fluids\/} {\bf 21}, 111701.

\bibitem[Duguet {\em et~al.\/}(2012)Duguet, Schlatter, Henningson \&
  Eckhardt]{Duguet2012}
Duguet, Y, Schlatter, P, Henningson, D~S \& Eckhardt, B 2012 {Self-sustained
  localized structures in a boundary-layer flow}. {\em Phys Rev Lett\/} {\bf
  108}, 044501.

\bibitem[Eckhardt \& Schneider(2008)]{Eckhardt2008}
Eckhardt, B \& Schneider, T~M 2008 {How does flow in a pipe become turbulent?}
  {\em The European Physical Journal B\/} {\bf 64}, 457--462.

\bibitem[Faisst \& Eckhardt(2003)]{Faisst2003}
Faisst, H \& Eckhardt, B 2003 {Traveling waves in pipe flow}. {\em Phys Rev
  Lett\/} {\bf 91}, 224502.

\bibitem[Fransson(2001)]{Fransson2001}
Fransson, J H~M 2001 {Investigations of the asymptotic suction boundary layer}.
  PhD thesis, KTH, Stockholm.

\bibitem[Gibson(2012)]{channelflow}
Gibson, J~F 2012 {{Channelflow}: {A} spectral {Navier-Stokes} simulator in
  {C}++}. {\em Tech. Rep.\/}. U. New Hampshire.

\bibitem[Gibson {\em et~al.\/}(2008)Gibson, Halcrow \&
  Cvitanovi\'{c}]{Gibson2008a}
Gibson, J~F, Halcrow, J \& Cvitanovi\'{c}, P 2008 {Equilibrium and
  traveling-wave solutions of plane Couette flow}. {\em J Fluid Mech\/} {\bf
  638}, 23.

\bibitem[Grossmann(2000)]{Grossmann2000}
Grossmann, S 2000 {The onset of shear flow turbulence}. {\em Rev Mod Phys\/}
  {\bf 72}, 603--618.

\bibitem[Halcrow {\em et~al.\/}(2009)Halcrow, Gibson, Cvitanovi\'{c} \&
  Viswanath]{Halcrow2009}
Halcrow, J, Gibson, J~F, Cvitanovi\'{c}, P \& Viswanath, D 2009 {Heteroclinic
  connections in plane {Couette} flow}. {\em J Fluid Mech\/} {\bf 621},
  365--376.

\bibitem[Hocking(1975)]{Hocking1975}
Hocking, L~M 1975 {Non-linear instability of the asymptotic suction velocity
  profile}. {\em Q J Mech Appl Math\/} {\bf 28}, 341.

\bibitem[Hof {\em et~al.\/}(2004)Hof, van Doorne, Westerweel, Nieuwstadt,
  Faisst, Eckhardt, Wedin, Kerswell \& Waleffe]{Hof2004}
Hof, B, van Doorne, C W~H, Westerweel, J, Nieuwstadt, F T~M, Faisst, H,
  Eckhardt, B, Wedin, H, Kerswell, R~R \& Waleffe, F 2004 {Experimental
  observation of nonlinear traveling waves in turbulent pipe flow}. {\em
  Science\/} {\bf 305}, 1594--1598.

\bibitem[Jeong \& Hussain(1995)]{Jeong1995}
Jeong, J \& Hussain, F 1995 {On the identification of a vortex}. {\em J Fluid
  Mech\/} {\bf 285}, 69--94.

\bibitem[Kawahara(2005)]{Kawahara2005a}
Kawahara, G 2005 {Laminarization of minimal plane Couette flow: Going beyond
  the basin of attraction of turbulence}. {\em Phys Fluids\/} {\bf 17}~(4),
  041702.

\bibitem[Kawahara \& Kida(2001)]{Kawahara2001}
Kawahara, G \& Kida, S 2001 {Periodic motion embedded in plane {Couette}
  turbulence: regeneration cycle and burst}. {\em J Fluid Mech\/} {\bf 449},
  291--300.

\bibitem[Khapko {\em et~al.\/}(2013)Khapko, Kreilos, Schlatter, Duguet,
  Eckhardt \& Henningson]{Khapko2013}
Khapko, T, Kreilos, T, Schlatter, P, Duguet, Y, Eckhardt, B \& Henningson, D~S
  2013 {Localized edge states in the asymptotic suction boundary layer}. {\em J
  Fluid Mech\/} {\bf 717}, R6.

\bibitem[Kreilos \& Eckhardt(2012)]{Kreilos2012}
Kreilos, T \& Eckhardt, B 2012 {Periodic orbits near onset of chaos in plane
  Couette flow}. {\em Chaos\/} {\bf 22}~(4), 047505.

\bibitem[Kreilos {\em et~al.\/}(2013)Kreilos, Veble, Schneider \&
  Eckhardt]{Kreilos2013}
Kreilos, T, Veble, G, Schneider, T~M \& Eckhardt, B 2013 {Edge states for the
  turbulence transition in the asymptotic suction boundary layer}. {\em Journal
  of Fluid Mechanics (accepted)\/} .

\bibitem[Lebovitz(2009)]{Lebovitz2009}
Lebovitz, N~R 2009 {Shear-flow transition: the basin boundary}. {\em
  Nonlinearity\/} {\bf 22}, 2645.

\bibitem[Mellibovsky {\em et~al.\/}(2009)Mellibovsky, Meseguer, Schneider \&
  Eckhardt]{Mellibovsky2009}
Mellibovsky, F, Meseguer, A, Schneider, T~M \& Eckhardt, B 2009 {Transition in
  Localized Pipe Flow Turbulence}. {\em Phys Rev Lett\/} {\bf 103}~(5), 1--4.

\bibitem[Nagata(1990)]{Nagata1990}
Nagata, M 1990 {Three-dimensional finite-amplitude solutions in plane {Couette}
  flow: bifurcation from infinity}. {\em J Fluid Mech\/} {\bf 217}, 519--527.

\bibitem[Schneider \& Eckhardt(2009)]{Schneider2009}
Schneider, T~M \& Eckhardt, B 2009 {Edge states intermediate between laminar
  and turbulent dynamics in pipe flow}. {\em Phil Trans R Soc A\/} {\bf
  367}~(1888), 577--87.

\bibitem[Schneider {\em et~al.\/}(2007{\natexlab{{\em a\/}}})Schneider,
  Eckhardt \& Vollmer]{Schneider2007a}
Schneider, T~M, Eckhardt, B \& Vollmer, J 2007{\natexlab{{\em a\/}}}
  {Statistical analysis of coherent structures in transitional pipe flow}. {\em
  Phys Rev E\/} {\bf 75}, 66313.

\bibitem[Schneider {\em et~al.\/}(2007{\natexlab{{\em b\/}}})Schneider,
  Eckhardt \& Yorke]{Schneider2007b}
Schneider, T~M, Eckhardt, B \& Yorke, J~A 2007{\natexlab{{\em b\/}}}
  {Turbulence transition and the edge of chaos in pipe flow}. {\em Phys Rev
  Lett\/} {\bf 99}, 34502.

\bibitem[Schneider {\em et~al.\/}(2010{\natexlab{{\em a\/}}})Schneider, Gibson
  \& Burke]{Schneider2010}
Schneider, T~M, Gibson, J~F \& Burke, J 2010{\natexlab{{\em a\/}}} {Snakes and
  ladders: localized solutions of plane {Couette} flow}. {\em Phys Rev Lett\/}
  {\bf 104}, 1--4.

\bibitem[Schneider {\em et~al.\/}(2010{\natexlab{{\em b\/}}})Schneider, Marinc
  \& Eckhardt]{Schneider2010a}
Schneider, T~M, Marinc, D \& Eckhardt, B 2010{\natexlab{{\em b\/}}} {Localized
  edge states nucleate turbulence in extended plane {Couette} cells}. {\em J
  Fluid Mech\/} {\bf 646}, 441.

\bibitem[Skufca {\em et~al.\/}(2006)Skufca, Yorke \& Eckhardt]{Skufca2006}
Skufca, J~D, Yorke, J~A \& Eckhardt, B 2006 {Edge of chaos in a parallel shear
  flow}. {\em Phys Rev Lett\/} {\bf 96}, 174101.

\bibitem[Toh \& Itano(2003)]{Toh2003}
Toh, S \& Itano, T 2003 {A periodic-like solution in channel flow}. {\em J
  Fluid Mech\/} {\bf 481}, 67--76.

\bibitem[Viswanath(2007)]{Viswanath2007}
Viswanath, D 2007 {Recurrent motions within plane {Couette} turbulence}. {\em J
  Fluid Mech\/} {\bf 580}, 339--358.

\bibitem[Vollmer {\em et~al.\/}(2009)Vollmer, Schneider \&
  Eckhardt]{Vollmer2009}
Vollmer, J, Schneider, T~M \& Eckhardt, B 2009 {Basin boundary, edge of chaos,
  and edge state in a two-dimensional model}. {\em New J Phys\/} {\bf 11},
  1--23.

\bibitem[Wang {\em et~al.\/}(2007)Wang, Gibson \& Waleffe]{Wang2007}
Wang, J, Gibson, J~F \& Waleffe, F 2007 {Lower branch coherent states in shear
  flows: Transition and control}. {\em Phys Rev Lett\/} {\bf 98}, 6--8.

\bibitem[Wedin \& Kerswell(2004)]{Wedin2004}
Wedin, H \& Kerswell, R~R 2004 {Exact coherent structures in pipe flow:
  travelling wave solutions}. {\em J Fluid Mech\/} pp. 1--42.

\bibitem[Willis \& Kerswell(2008)]{Willis2008}
Willis, A~P \& Kerswell, R~R 2008 {Turbulent dynamics of pipe flow captured in
  a reduced model: puff relaminarization and localized ‘edge’ states}. {\em
  J Fluid Mech\/} {\bf 619}~(2009), 213.

\end{thebibliography}

\end{document}